\documentstyle[conf_iap,psfig,10pt]{article}
\begin{document}
\heading{The ESO-Sculptor Survey: Galaxy Populations and Luminosity
Function at $z \le 0.5$} 
\par\medskip\noindent
\author{Gaspar Galaz$^1$ and Val\'erie de Lapparent$^2$}
\address{Observatories of the Carnegie Institution of Washington. Las
Campanas Observatory. Casilla 601, La Serena, Chile (gaspar@ociw.edu)}
\address{Institut d'Astrophysique de Paris, CNRS. 98bis, Boulevard Arago,
75014, Paris, France (lapparen@iap.fr)}

\begin{abstract}
We briefly show results on the redshift and space distribution of 
field galaxies with different spectral types in the ESO-Sculptor 
survey (ESS). We also show results on the ESS galaxy 
luminosity function. 
%
\end{abstract}
\section{Introduction: The ESO-Sculptor Survey}

The ESO-Sculptor Survey \cite{lapparent97} provides photometric 
and spectroscopic data of galaxies from a narrow, but deep, patch of the
sky. 
Data were gathered at La Silla Observatory (ESO), using both the 3.6m
telescope and the NTT. The photometric 
catalogue covers a continuous strip of 1.53$^\circ$ (R.A.) $\times$
0.24$^\circ$ (DEC.) $\sim 0.37$ deg$^2$ in the Sculptor constellation, 
with center (J2000) $\sim 0^h$ 21$^m$, $-30^\circ$, and 17$^\circ$ from 
the south galactic pole. A detailed description of the construction and
reduction of the photometric catalogue can be found in \cite{arnouts97}.

The spectroscopic catalogue contains the spectra and redshifts of
$\sim 700$ galaxies with R = 20.5, obtained 
using multi-slit spectroscopy. 
Details on the reduction of the spectroscopic data are in \cite{bellanger95}. 
The ESS has already provided
some results on our view of the distribution of galaxies at large-scale to 
$z \sim 0.5$ \cite{bellanger95b}.
Here we describe recent 
results on the proportion and distribution of the different types of 
galaxies, provided by a spectral classification approach
based on the principal component analysis (PCA) \cite{murtagh87}. 
This spectral classification is fundamental to have 
insights on the distribution of different galaxy types which populate
the large-scale structures up to $z \sim 0.5$. Furthermore, it 
allows to compute K-corrections which in turn allow 
to have rest-frame magnitudes to construct the galaxy luminosity
function (LF). These main results follow.

\section{Galaxy Populations}

Understanding the formation and evolution of galaxies \cite{baugh96} requires 
a detailed knowledge of the different galaxy populations as a 
function of redshift and in terms of the local density. In 
order to obtain an unbiased description of the different 
galaxy populations for the ESS, we have used a spectral classification
approach which is more tightly related to the underlying 
physics and to the stellar populations which characterize the galaxies, than
would be a morphological classification. Morphological classifications
also have the inconvenient of being filter and redshift dependent.
The PCA spectral classification for the ESS was made using 
only spectra taken during spectro-photometric
nights, because PCA galaxy spectral types are partly determined
by the shape of the continuum \cite{connolly95}. A sub-sample of 330
spectra were analyzed in the spectral range $\lambda = 3700 - 5250$ \AA, 
where the best compromise in terms of spectral coverage and
number of spectra is obtained. The PCA allows to 
generate a {\em spectral classification plane} for galaxies, which 
provides a continuous spectral sequence for the ESS (labeled from I [early
types] to VI [late types]). All details on 
the ESS spectral classification are provided in \cite{galaz98}.

\begin{figure}[ht]
\centerline{\hbox{
\psfig{figure=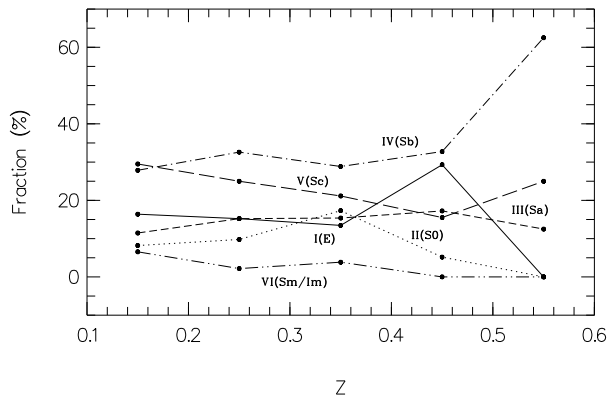,height=6.0cm,angle=-90}}}
\caption[]{The redshift distribution of the different PCA spectral types
in the ESS. Spectral types are labeled in roman. Few galaxies 
are detected at $z > 0.5$. See text for details.}
\label{types_z_ra}
\end{figure}

The analysis of the ESS spectral types as a function of redshift, 
shows that there is {\em no} significant evolution 
of different spectral types up to $z \sim 0.5$.  
Except for an excess of early-type galaxies at $z \sim 0.43$ 
(caused by a large galaxy concentration), 
the distribution per redshift is nearly 
uniform for each spectral type as a function of redshift 
(see Figure \ref{types_z_ra}). 
Another important result is related to the morphology-density relationship
\cite{dressler97}.
The distribution of different spectral types as a function of the 
local density shows that most of the early-type galaxies ($\sim 85\%$)
are associated to galaxy groups. 
Late-type galaxies, on the other hand, occupy
nearly uniformly both the low and high density regions.

\section{The ESS Luminosity Function}

The projection of spectrophotometric templates onto the ESS classification 
plane \cite{galaz98} allows to compute K-corrections based on the 
spectral classification. 
This analysis requires the use of synthetic spectra which allow to
``extrapolate'' the ESS spectra in the standard filters when these are
not included in the spectral range of the observed spectra.
We have used the PEGASE models \cite{fioc97} using a star formation 
rate proportional to the amount of gas, and solar metallicity. Ages range 
from 0.1 Gyr to 16 Gyr. For each ESS spectrum, 
the closest projected model onto the classification plane (with the same
wavelength interval as for the ESS spectrum) is taken as its most
similar model spectrum. Synthetic photometry is done onto the model
spectra using the B$_j$, V$_j$ and R$_c$ ESS
passbands and K-corrections are computed following their definition 
\cite{oke68}. 
Although this approach depends on the model which is used, the
source of difference from one model to another can be controled.
These K-corrections are more reliable than
those obtained by the standard approach of using morphological 
classification and making strong hypotheses on the relationship between the 
spectral energy distributions (SEDs) and the morphological types. 
Measuring the K-corrections is one of the fundamental steps 
to construct the galaxy luminosity function (LF).  
   
The ESS LF was calculated using standard maximum likelihood techniques,
including a non-parameterized (EEP method, \cite{eep88}) and 
a parameterized functional form (using the STY method, \cite{sty79}). 
For the latter, we adopt the Schechter function. 
Maximum likelihood methods ensure that the LF is not biased by the 
large-scale inhomogeneities present in the catalogue.  
\begin{figure}[ht]
\centerline{\vbox{
\psfig{figure=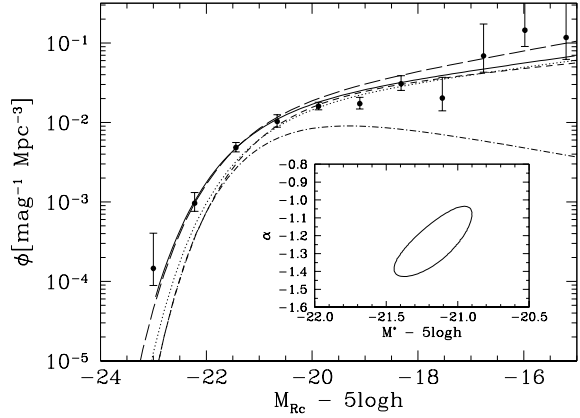,height=6.5cm,angle=-90}}}
\caption[]{The ESS galaxy luminosity function (LF) in the R$_c$ (Cousins)
filter. The bold line represents the best-fit Schechter LF function 
(see text for details). 
LFs for other surveys are also shown using their
Schechter fit: the ESP, dotted line \cite{zucca97}; 
the CNOC1, long-dashed \cite{lin97}; 
the Century Survey, short-dashed \cite{geller97}; the LCRS, dot-dashed
\cite{lin96}. The inset 
shows the 1-sigma error ellipse for the ESS Schechter LF. Points with error
bars represent the ESS non-parameterized (EEP) LF.} 
\label{lf_r}
\end{figure}
The best-fit Schechter function for the ESS LF in R$_c$ 
(M$^* = -21.15 \pm 0.19 + 5\log h$, $\alpha = -1.23 \pm 0.13$ and $\phi^* = 
0.0203 \pm 0.008 h^3$ Mpc$^{-3}$ (q$_0 = 0.5$, H$_0 = 100$ km 
s$^{-1}$ Mpc$^{-1}$) is shown as a bold line in Figure \ref{lf_r}. 
Symbols represent the EEP non parameterized LF. The 1-sigma error
ellipse for the Schechter parameters is shown in the inset. 
The R$_c$ band LF was constructed from 327 ESS galaxies
with PCA spectral types, and corrected for incompleteness effects
following the method described in \cite{zucca94}. 
Figure 2 also shows the LF of other 
surveys, transformed into the R$_c$ band. The LF for the ESS is in
good agreement 
with the LF for the CNOC1 survey (\cite{lin97}, long-dashed line), 
the ESO-Slice project (hereafter ESP, \cite{zucca97}; dotted line), and the Century 
survey (\cite{geller97}, short-dashed line).
The LF for the LCRS (\cite{lin96}, dash-dotted line) is
different from the other LFs, in particular at the faint end. Selection effects in
that survey are probably responsible for the lost of a large number of 
(faint) 
blue galaxies \cite{lin96}. Like in other surveys (in particular the ESP 
and the CNOC1 survey), we note that the faint end of the LF 
follows an exponential form for magnitudes fainter than M$_{Rc} = -18$. 
Most of the galaxies which occupy this magnitude
range are galaxies with spectral types later than IV 
(see Figure \ref{types_z_ra}).
82\% of these galaxies show evidence for present star formation, as measured
by the [OII]3727\AA\ emission
line. More than 80\% of the galaxies which have M$_{Rc} <$ M$^*$ have 
spectral types earlier than III, and very few ($<$2\%) show 
evidence for star formation.

\acknowledgements{GG acknowledge Institut d'Astrophysique de Paris and 
Carnegie Observatories support for this work.}


%
\vfill
\end{document}